\documentclass[%
 reprint,
superscriptaddress,
 amsmath,amssymb,
 aps,
]{revtex4-2}
\usepackage{ulem}
\usepackage[utf8]{inputenc}
\usepackage{lmodern} 
\usepackage{microtype}
\usepackage{natbib}
\setcitestyle{numbers,square,sort&compress}
\usepackage{amsmath,amssymb,amsfonts}
\usepackage{algorithmic}
\usepackage{graphicx}
\usepackage{textcomp}
\usepackage{lipsum,mathtools,cuted}
\usepackage{xcolor}

\usepackage{float}
\usepackage{epsfig}
\usepackage{epstopdf}
\usepackage{subfigure,color}
\usepackage{dcolumn}
\usepackage{bm}
\usepackage{lineno}
\usepackage{accents}
\usepackage{multirow}
\usepackage[usestackEOL]{stackengine}
\usepackage{tabularx}
\usepackage{makecell}
\usepackage{dcolumn}
\usepackage{bm}
\usepackage{booktabs}
\UseRawInputEncoding

\usepackage{mdframed}
\begin{document}
\title{Control of the local and non-local electromagnetic response in all-dielectric reconfigurable metasurfaces}

\author{L. M. M\'{a}\~{n}ez-Espina}
\affiliation{
Nanophotonics Technology Center, Universitat Polit\`ecnica de Val\`encia, Val\`encia 46022, Spain
}
\email{lmmaeesp@upv.es}
 
 \author{A. D\'iaz-Rubio}
\affiliation{
Nanophotonics Technology Center, Universitat Polit\`ecnica de Val\`encia, Val\`encia 46022, Spain
}

\date{\today}

\begin{abstract}
Reconfigurable metasurfaces are potent platforms to control the propagation properties of light dynamically. Among different reconfiguration mechanisms available at optical frequencies, using non-volatile phase change materials is one of the most prominent. Tuning the refractive index of these materials, and thus changing the electromagnetic response of the device, can be achieved by external modulation that enables the transition between their amorphous and crystalline structural states. In this work, we study a structure that exploits these materials to fully control the electromagnetic response, i.e., electric, magnetic, and electromagnetic coupling. Powered by this controllability, we present how the metasurface can be designed to perform beam steering and switchable one-way absorption applications. In addition, we demonstrate that the same platform can be designed to work in a nonlocal regime to implement a more efficient way of controlling light direction.
\end{abstract}

\keywords{All-dielectric metasurfaces, non-local metasurfaces, phase change materials, reconfigurable metasurfaces.}

\maketitle

\section{Introduction}

Since their inception, metasurfaces have gained significant attention in photonics due to their ability to control and manipulate electromagnetic (EM)  waves at subwavelength scales~\cite{yu_flat_2014,Kildishev_2013_planar,yu_light_2011}. These structures are comprised of two-dimensional arrays of subwavelength resonators that enhance light-matter interaction, opening up new possibilities for compact and efficient photonic devices with versatile functionalities, from polarization vortex creation~\cite{ahmed2022optical} to quantum optical applications~\cite{li2024metasurface,solntsev_metasurfaces_2021}. The main advantage of these structures is the capacity to engineer local and non-local effects, offering full control over light amplitude, phase, and polarization~\cite{asadchy2018bianisotropic,shastri2023nonlocal}.

Although metasurfaces have demonstrated the capacity to manipulate light and perform exotic transformations of waves in the entire EM spectrum, in most cases, their response is fixed by design. During the last decade, the concept of tunable, or reconfigurable, metasurfaces was introduced~\cite{Cui_2014_coding}. The broad concept is to use a variable parameter, or parameters, to change the EM response of the devices~\cite{gu2023reconfigurable, saifullah2022recent,badloe_tunable_2021}. Metasurfaces working at optical frequencies usually rely on changing the optical properties of their constituent materials by employing an external excitation. One of the most relevant reconfiguration mechanisms is using phase change materials (PCMs) that produce a change in the refractive index when transitioning from an amorphous state to a crystalline state~\cite{de_galarreta_nonvolatile_2018,yin_beam_2017,kaplan_dynamically_2015,zou_phase_2014,leitis_alldielectric_2020,liu_phase-change_2024}. Different external modulation mechanisms can control these transitions, usually by applying heat or optical pumping~\cite{Review_PCM}.

\begin{figure}[b]
    \centering
    \includegraphics[width=0.9\columnwidth]{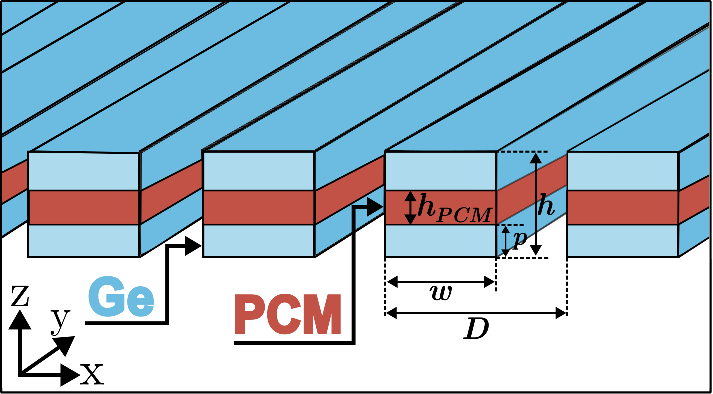}
    \caption{Metasurface representation and geometrical parameters. Infinite pillars in the $y$-direction with rectangular cross-section defined by its height, $h$, and width $w$. The periodicity is defined as $D$. The PCM slab is introduced at the center of the bars and has height $h_{\rm PCM}$. The parameter $p$ determines the position of the PCM layer.}
    \label{fig:1_Graphical_design}
\end{figure}

In this work, we focus on designing an EM reconfigurable metasurface using PCMs and their capability to fully control the EM response, i.e., electric and magnetic response and EM coupling. Specifically, we will target beam steering and absorption applications with the same metasurface topology as test case scenarios. The proposed metasurface is composed of a one-dimensional periodically spaced array of germanium nanobars with a PCM slab with refractive index, $n_{\rm PCM}$, inserted in a sandwich configuration with height $h_{\rm PCM}$, as shown in Fig.~\ref{fig:1_Graphical_design}. The position and the thickness of the PCM slab will be adapted depending on the application under study. The parameter $p$ determines the positions of the PCM layer. The meta-atoms are horizontally placed bars with rectangular cross-sections, considered infinite in the $y$-direction. The total height of the bars is defined as $h$, their width as $w$, and the one-dimensional periodicity of the structure is defined by the parameter $D$. 

For beam steering applications~\cite{wu_dynamic_2019,shirmanesh_electro-optically_2020,Zhang2021,Chu2016,Kim2022,berini_optical_2022,berini_optical_2022}, we demonstrate that the metasurface can produce a linear spatial distribution of phase with near-unity transmission efficiency, capable of redirecting the incoming energy to the first diffraction order. Following the generalized law of refraction~\cite{yu_light_2011} and, therefore, exploiting the local response of the structure, we demonstrate the capacity to control the direction of wave propagation in transmission. With the same configuration, the structure can be tuned to act as a retroreflector, exploiting the strong non-local response of spatially varying metasurface that couples the incoming wave to higher diffraction orders by engineering the near fields.

For absorption applications, we propose a different configuration that breaks mirror symmetry in the wave propagation direction and allows for controlling EM coupling. This type of weak-non-local behavior allows asymmetrical control of the reflection of the metasurface when it is illuminated from opposite directions~\cite{asadchy2018bianisotropic}. Our results show how we can obtain a tunable asymmetric absorber by exploiting the losses in the PCM materials. Asymmetric absorption is obtained while achieving out-of-band transparency. In contrast with most of the previous approaches to absorption tunability that used a ground plane~\cite{yao_electrically_2014,Han_220_Absorber0,MAURYA_2024_Tunable}, the proposed design does not use such a structure, allowing transparency out of the resonance.

\section{Electric and magnetic response tunability}

\begin{figure}[t]
    \centering
    \includegraphics[width=1\columnwidth]{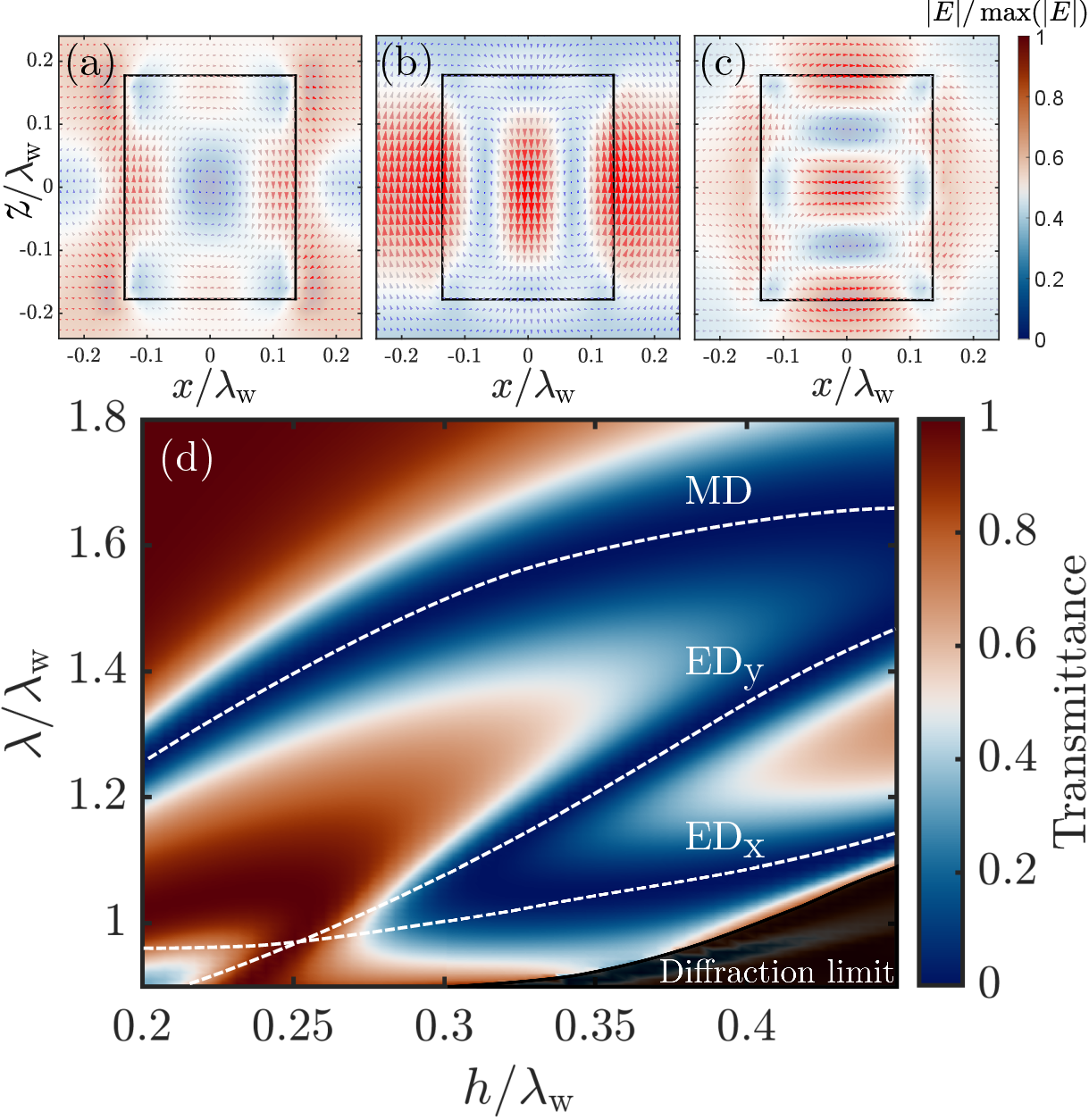}
    \caption{(a-c) Electric field intensity of the first three dipole resonant modes. Arrows show the electric field direction. (d) Colormap of the transmittance intensity. The metasurface is illuminated with a plane wave at an angle of incidence of 45$^{\circ}$ with TM polarization. The ED resonances cross approximately when $h=0.25\;\lambda_{\rm W}$, allowing full transmission. The results for combinations of wavelengths and heights above the diffraction limit have been shaded. The geometrical parameters for figures (a-d) are $D=0.52\;\lambda_{\rm W}$, $w=0.27\;\lambda_{\rm W}$, $n_{\rm PCM}=n_{\rm Ge}$=4 }
    \label{fig:2_Resonances_and_Huygens_pair}
\end{figure}

Due to the $x$-$y$ plane geometrical anisotropy, the structure shown Fig.~\ref{fig:1_Graphical_design} will behave differently when illuminated by plane waves linearly polarized with orthogonal polarizations. A full-wave simulation is performed to study the resonant behavior of the metasurface when it is illuminated with a linearly polarized wave with TM polarization, $H_y$ is the only component of the magnetic field different from zero, with a wavevector contained in the $z$-$x$ plane that is tilted $45^{\circ}$ with respect to the $z$-axis (i.e., the wave vector is contained in the plane perpendicular to the bars and the electric field of the wave is contained in such plane). The first three induced EM moments are a magnetic dipole in $y$ ($\rm MD$), an electric dipole in $z$ ($\mathrm{ED}_z$), and an electric dipole in $x$ ($\mathrm{ED}_x$). Figure~\ref{fig:2_Resonances_and_Huygens_pair} (a)-(c) represents the electric field of the resonant modes supported in the structure calculated with an eigenmode analysis with $h/\lambda_{\rm W}=0.35$. The parameter $\lambda_{\rm w}$ represents the wavelength at the operation frequency, the desired wavelength where the device must work. The first resonance, shown in Fig.~\ref{fig:2_Resonances_and_Huygens_pair} (a) and located at $\lambda/\lambda_{\rm w}=1.6$ corresponds to a $y$-directed magnetic dipole and it is created by the circulation of the displacement currents around the perimeter of the rectangular cross-section.  The second resonance is shown in Fig.~\ref{fig:2_Resonances_and_Huygens_pair} (b) and corresponds with an electric dipole in the $z$-direction located at $\lambda/\lambda_{\rm w}=1.2$. The last resonance is placed at $\lambda/\lambda_{\rm w}=1.05$ and depicts a $x$-oriented electric dipole as it is shown in Fig.~\ref{fig:2_Resonances_and_Huygens_pair} (c).

Figure \ref{fig:2_Resonances_and_Huygens_pair} (d) shows the transmittance, defined as the absolute value of the transmission coefficient squared. The colormap depicts the dependency on the height of the nanobars when the period of the structure is fixed to $D=0.52\;\lambda_{\rm W}$, the width $w=0.27\;\lambda_{\rm W}$ and the refractive index of the PCM is $n_{\rm PCM}=n_{\rm Ge}=4$. Notice that in this configuration, $n_{\rm PCM}=n_{\rm Ge}$ and the structure can be modeled as a homogeneous nanobar made of germanium. The resonant modes produce zero transmission, and their dependency on the parameter $h$ is marked with white dashed lines. As expected, an increment in the height produces a displacement of the three resonant modes towards higher wavelengths. From this analysis, one can see that the two electric modes,  $\mathrm{ED}_x$ and  $\mathrm{ED}_z$, change differently with the parameter $h$, allowing the overlap of both resonances at the same frequency. As it was demonstrated, in~\cite{kwon2018transmission}, zero backscattering and full control of the phase in transmission could be obtained by bringing together the symmetric (respect to the $z$-axis) $\mathrm{ED}_x$ resonance and the antisymmetric $\mathrm{ED}_{z}$. Under this condition, the structure forms a Huygens' pair that can be used to control the direction of propagation of the transmitted EM waves. 

\begin{figure}[t]
    \centering
    \includegraphics[width=1\columnwidth]{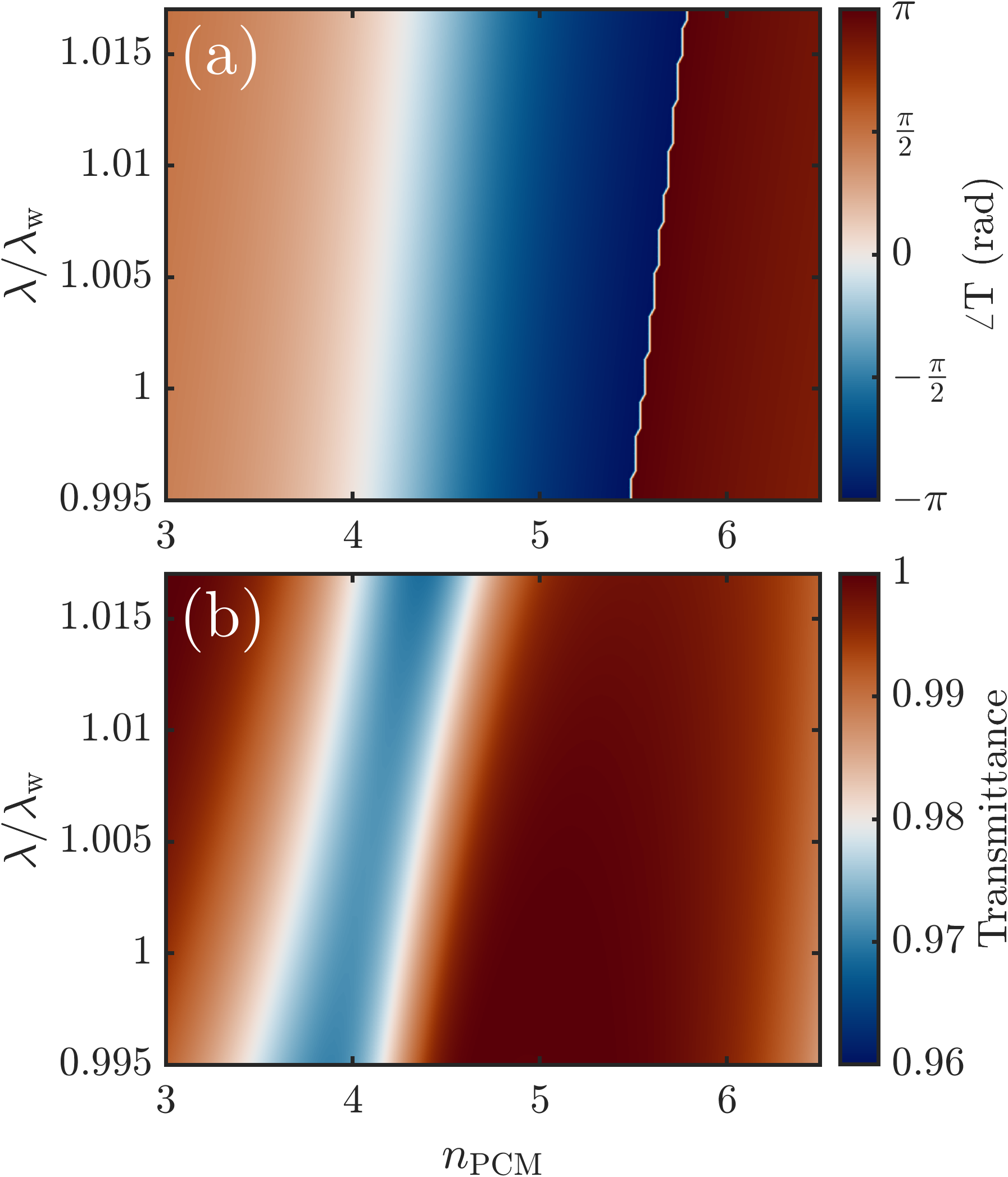}
    \caption{Phase and transmittance shown in (a) and (b), respectively, of a metasurface with geometrical parameters: $D=0.52\;\lambda_{\rm W}$, $w=0.27\;\lambda_{\rm W}$, $h=0.25\;\lambda_{\rm W}$, $p=h/3$,$n_{\rm PCM}=n_{\rm Ge}=4$, and $h_{\rm PCM}$=$h/3$. These results are obtained under TM polarized plane wave illumination at $45^{\circ}$. }
    \label{fig:3_Phase_and_transmittance}
\end{figure}

Once the resonance pair is constructed, the geometrical parameters and $\lambda_{\rm W}$ are left fixed, and we can study the effect of changing the refractive index in the PCM layer. The height of the PCM is set to be $h_{\rm PCM}=h/3$ and $p=h/3$. In Fig.~\ref{fig:3_Phase_and_transmittance}, the phase and transmittance of the metasurface are represented for the range of $n_{\rm PCM}$ between 3 and 6.5. Transmittance is near unity for all the refractive index values, Fig.~\ref{fig:3_Phase_and_transmittance}(b). Beyond the limit in refractive index ($n_{\rm PCM}>6.5$), the PCM layer is so electrically large that the displacement field produces a resonance inside this layer, detuning the Huygens pair. When ($3<n_{\rm PCM}<6.5$), the change of the refractive index in the central slab shifts each ED resonance by roughly the same amount so the pair is close enough to create high transmittance while changing the phase drastically. The phase coverage in the represented range is almost complete, Fig.~\ref{fig:3_Phase_and_transmittance}(a). It is important to notice that the chosen configuration, a central layer of PCM sandwiched between two dielectric germanium layers, is symmetric in the $z$-direction, neglecting EM coupling and maintaining high transmission. The resonance pair is stronger in the center of the rectangular bar; therefore, the refractive index change has more effect on both ED resonances and can achieve a great change in the phase of the transmitted wave. Furthermore, the geometry prevents polarization conversion, as shown in Ref.~\cite{kwon2018transmission}.

The presented tunability properties can be used to control the direction of the propagating waves by implementing different phase profiles. Thus, the meta-atoms are arranged in groups of $N$ members to create a periodic structure with period $D_{\rm N}=ND$. The ensemble of meta-atoms is going to be referred to as a meta-cell. Notice that the period $D_{\rm N}$, with $N>1$, is large enough to allow Floquet modes of higher order to propagate, and the energy of the incident wave can be coupled to the Floquet modes using two different mechanisms. On the one hand, using a \textit{local approach}, the proposed topology can implement a phase gradient in transmission that can control the direction of transmitted waves, behaving as a diffraction grating. On the other hand, using a \textit{non-local approach}, the topology allows the tailoring of the evanescent field created by the structure and channel of the energy into the desired Floquet mode. In the following subsections, we show examples of metasurfaces capable of controlling the direction of propagation using both physical mechanisms. 

\subsection{Beamsteering using local gradient approach}

 \begin{figure}[t]
    \centering
    \includegraphics[width=1\columnwidth]{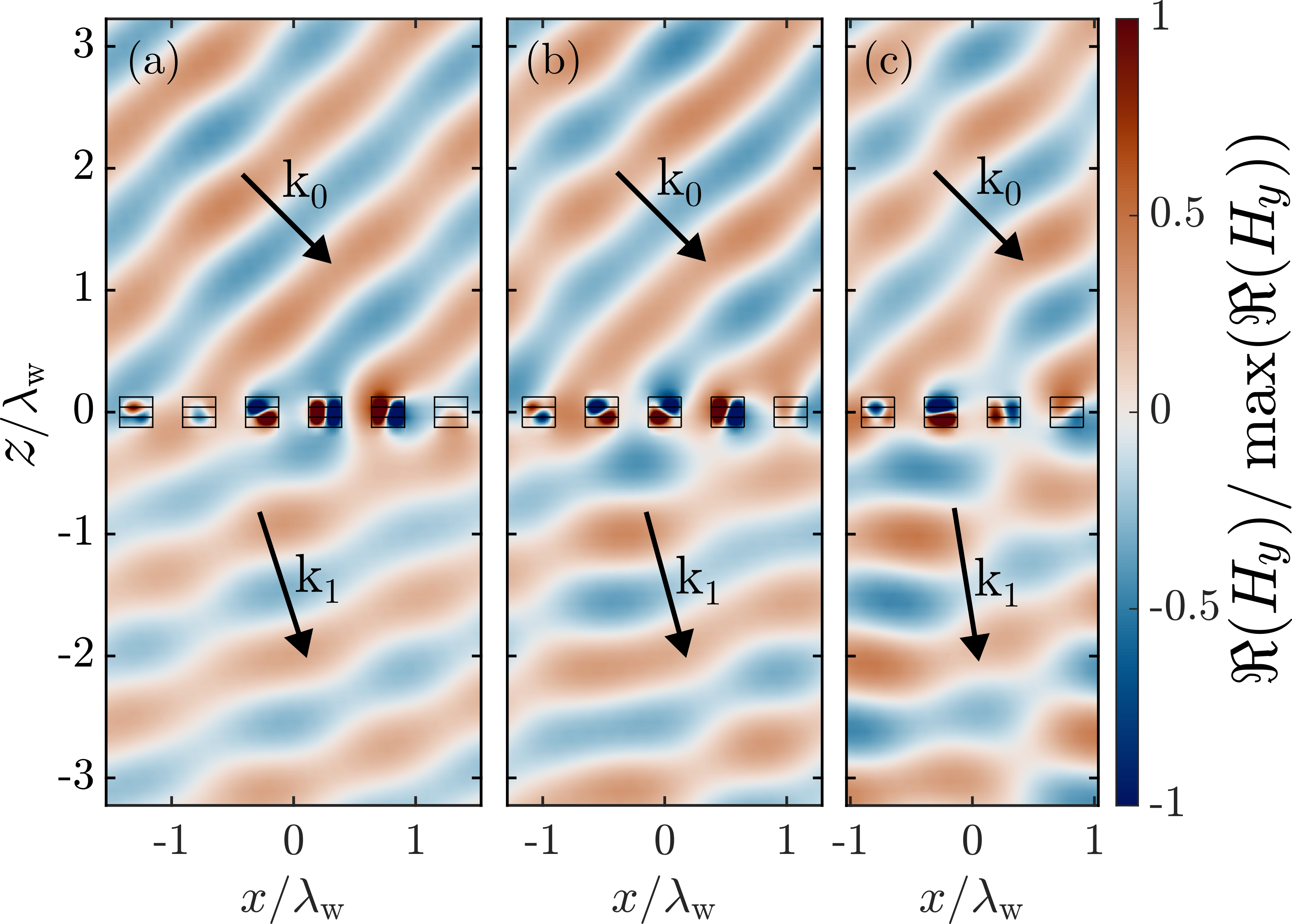}
    \caption{Normalized real part of $H_y$ for various meta-cell configurations. TM linearly polarized plane-wave illumination at 45$^{\circ}$ is represented by the wavevector $\rm k_0$. The outgoing wavevector for the first diffraction order in transmission is also shown as $\rm k_1$. The geometrical parameters for the metasurface are the same as used in Fig.~\ref{fig:3_Phase_and_transmittance}. The refractive index for PCM in the resonators, labeled from left to right, are in each case, with notation $\overline{n}=\left[n_1,n_2,n_3,...,n_N\right]$: (a) $\overline{n}=\left[6.6,5.1,4.6,4.3,3.8,2.7\right]$; (b) $\overline{n}=\left[6.1,4.9,4.4,3.9,2.7\right]$; (c)  $\overline{n}=\left[5.6,4.6,4.1,2.7\right]$. The angles for the outcoming waves are  $22.5^{\circ}$, $18.6^{\circ}$, and $12.8^{\circ}$ for the cases (a), (b) and (c), respectively.}
    \label{fig:4_6_5_4_elements_phase_gradient}
\end{figure}

First, we start designing a phase gradient metasurface based on a local approach that controls the direction of propagation of transmitted waves. As it was proposed in~\cite{yu_light_2011}, the direction of transmitted waves can be controlled by applying the following phase gradient
\begin{equation}\label{eq:general_diffraction}
    {\rm sin}(\theta_{\rm t}) n_{\rm t}-{\rm sin}(\theta_{\rm i})n_{\rm i}=\dfrac{\lambda_0}{2\pi}\dfrac{d\Phi(x)}{dx},
\end{equation}
where $\theta_{\rm t}$, $n_{\rm t}$, $\theta_{\rm i}$, and $n_{\rm i}$ are the angles and refractive indexes of the second medium and the first medium, respectively. The function $\Phi(x)$ defines the phase discontinuity along the metasurface, and $\lambda_0$ is the vacuum wavelength. Such phase discontinuity can be implemented using resonator elements with different relative phases in a linear fashion. In practice, the phase discontinuity is inherently noncontinuous and, therefore, is an approximation that reduces the efficiency of this approach. When the derivative in Eq.~(\ref{eq:general_diffraction}) is a constant, the generalized law of refraction can also be interpreted as a recipe to obtain high energy coupling to the first Floquet mode~\cite{larouche_reconciliation_2012,schake_examining_2022} (first diffraction mode) in transmission.

The previous result allows the design of a discretized linear phase gradient in the $x$-direction by selecting a group of $N$ elements, creating a meta-cell with periodicity $D_N$, and choosing an adequate refractive index for each element to cover the $2\pi$ range in phase. The created gradient in phase will be a discrete approximation to the continuous function with a gradient value of
\begin{equation}\label{eq:Phase_gradient}
    \dfrac{d\Phi (x)}{dx}=\dfrac{2\pi}{D_N}.
\end{equation}

At a fixed incoming angle, $\theta_{\rm i}$, the selected number of meta-atoms in the meta-cell will impact the value of Eq.~(\ref{eq:Phase_gradient}), and therefore the transmitted angle, $\theta_{\rm t}$, in Eq.~(\ref{eq:general_diffraction}). By changing the number of elements per meta-cell and the individual configuration of each meta-atom, the direction of propagation of the transmitted waves can be changed. It is important to notice that, geometrically, the metasurface is equal in every situation, and only the refractive index of the PCM is what changes. Moreover, with this configuration, the directions for the outcoming waves are discrete, as the number of diffraction modes is finite and discrete. Continuous beam steering is, therefore, not allowed. Such operation can be achieved when the period is changed in a continuous fashion~\cite{CONTINUOUS}.

The results for the six-, five-, and four-element meta-cells are shown in Fig.~\ref{fig:4_6_5_4_elements_phase_gradient}. As the wavevector of the incoming illumination is contained in the perpendicular plane and there is no polarization conversion, $H_y$ is the only non-zero component of the magnetic field. The normalized real part of $H_y$ is represented in Fig.~\ref{fig:4_6_5_4_elements_phase_gradient} and shows how the incident wave at a 45$^{\circ}$ angle, marked by the black arrow that represents the wavevector $\mathrm{k_0}$, is converted into a transmitted wave with wavevector $\mathrm{k_1}$ at different angles for each meta-cell. The highest field intensity is found inside the resonators, which are out of phase to create the phase gradient and have roughly the same scattering amplitude. As the number of elements decreases, the wave in the lower semi-plane is less pure because higher diffraction orders are excited and interfere with the first diffraction order.

Using the local approach, the maximum energy coupling is bounded and depends on the incident and transmitted angles. Several studies have calculated the theoretical maximum energy coupling from the incoming wave to the first Floquet mode that can be achieved following this local approach~\cite{asadchy_perfect_2016,diaz-rubio_generalized_2017}. In addition to this fundamental limitation, another important assumption that is made in this approach is considering the almost local behavior of the device that is going to replicate the discontinuity. In practice, each resonator will inevitably interact with its neighbors, and the response of the whole system will not be possible to interpret by only taking into account the pointwise response. This is referred to in the literature as nonlocality~\cite{overvig_diffractive_2022}, and, as we will show in the next section, it can be used to channel energy in the desired direction with high efficiency. However, although nonlocality is always present in the system response, it can be minimized if the size of the meta-atoms is small and the geometry between adjacent meta-atoms does not drastically change. These two aspects limit the response of the metasurface and, as shown in Fig.~\ref{fig:4_6_5_4_elements_phase_gradient}, produce undesired scattering in other directions. 

\begin{figure}[t]
    \centering
    \includegraphics[width=1\columnwidth]{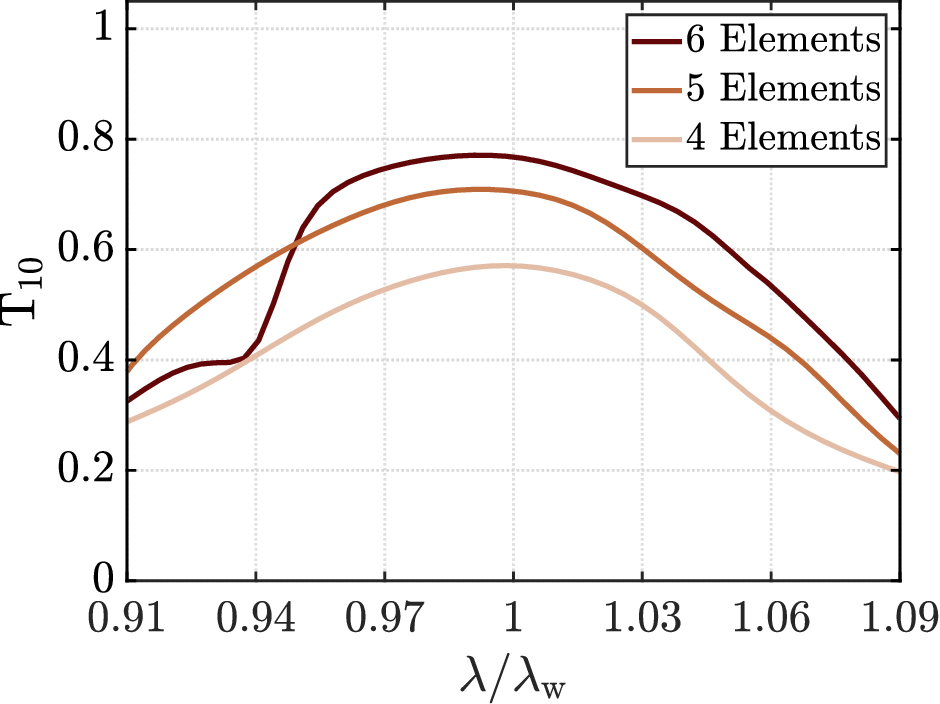}
    \caption{Transmittance from the zero-order of diffraction to the first order of diffraction for the three arrangements shown in Fig.4. The maximum transmittance is obtained near the working wavelength. }
    \label{fig:5_first_order_efficiencies}
\end{figure}

Further analysis of the efficiency of the structure is shown in Fig.~\ref{fig:5_first_order_efficiencies} where the power transmitted into the desired direction is represented for the three implementations shown in Fig.~\ref{fig:4_6_5_4_elements_phase_gradient}. In this study, $\mathrm{T_{10}}$ is the transmittance, the squared norm of the transmission coefficient, between the 0-order channel, which is the incident wave, and the first order of diffraction in transmission. The transmittance values for other diffraction orders are not plotted as they are minor contributors. The values of this transmittance can be understood as the energy coupling between incident light and the first order of diffraction. It is apparent from Fig.~\ref{fig:4_6_5_4_elements_phase_gradient} and Fig.~\ref{fig:5_first_order_efficiencies} that the highest efficiency is obtained in the 6-element scenario, reaching $77\%$ coupling efficiency. The result is due to the better discretization of $\Phi (x)$. More elements make a more faithful approximation to a linear phase profile than fewer elements. The results shown serve as a proof of concept for a reconfigurable platform capable of redirecting the incident wave to different angles by selecting different meta-cells with the simple geometry shown in Fig.~\ref{fig:1_Graphical_design}. 

\subsection{Beamsteering using strong non-local approach}

The local assumption was useful in the last section to create a geometrical dispersive gradient metasurface capable of redirecting the incoming wave into different angles by changing the meta-cell used. However, different capabilities can be achieved by engineering the interactions between meta-atoms via near-field excitation. The response of such a system is usually referred to as nonlocal~\cite{overvig_diffractive_2022,overvig_multifunctional_2020,chai_emerging_2023,shastri_nonlocal_2023}. The behavior of nonlocal metasurfaces, by definition, cannot be explained by the response of each resonator alone. The EM response of highly nonlocal metasurfaces is a collective effect that is hard to describe analytically. Many efforts have been made, and some recent publications have focused on explaining the underlying physics involved in nonlocal metasurfaces using phenomenological models~\cite{overvig_spatio-temporal_2024}. Despite this complication, powered by the computational capabilities of current EM simulators, it is possible to use optimization techniques to obtain nonlocal responses with a certain performance that exceeds the efficiency limits of the local approach or enables different functionalities. Indeed, high coupling to different Floquet modes can be achieved by mechanisms not apparent in the study of the resonances at play alone and resulting from complex near-field interactions. 

\begin{figure}[t]
    \centering
    \includegraphics[width=1\columnwidth]{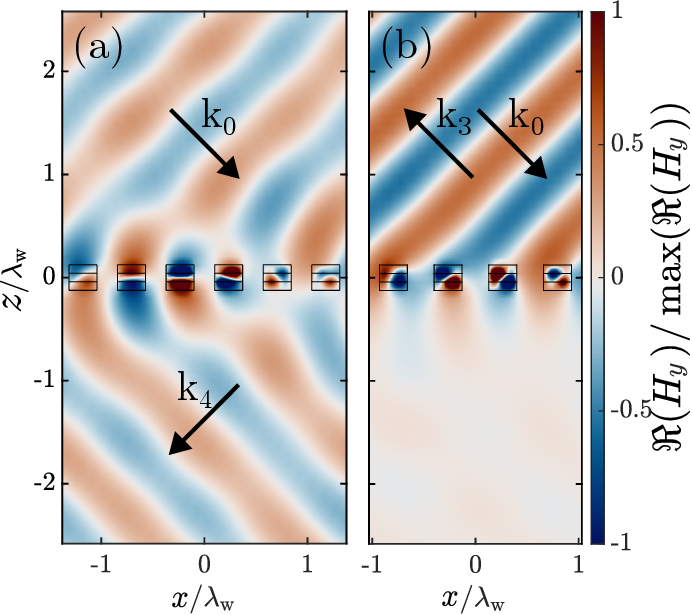}
    \caption{Field colormap of the $y$ component of the magnetic field. Results obtained with plane wave TM polarized illumination at $45^{\circ}$, wavevector $\rm k_0$. (a) The principal excited diffraction order is the fourth in transmission with an efficiency of $91.2\%$. The outgoing wave makes an angle of -$45^{\circ}$ at wavelength $\lambda_w$. The parameters of the structure are $D=0.47\;\lambda_{\rm W}$, $N=6$ and $\overline{n}=\left[2.82, 3.45, 4.16, 4.51, 5.04, 6.50\right]$. (b) The third mode of refraction in reflectance is excited with an efficiency of $98.9\%$ at $\lambda_w$. The outcoming wavevector makes an angle of -$45^{\circ}$ with the $z$-axis. The structure has parameters $D=0.53\;\lambda_{\rm W}$, $N=4$ and $\overline{n}=\left[3.12,4.19,4.66,5.90\right]$. The rest of the geometrical parameters for (a) and (b) are the same as in Fig.~\ref{fig:3_Phase_and_transmittance}.}
    \label{fig:6_Non_local_responses}
\end{figure}

The design approach for a non-local metasurface that controls the direction of propagation starts by defining the total period, $D_N$, that allows the energy to propagate into the desired direction. With the proposed topology, the period will be multiple of $D$. In this scenario, because the energy coupling is not dominated by an induced phase gradient along the metasurface but for the excitation of evanescent auxiliary fields, the energy can be coupled to any Floquet mode and not only to the first mode as in the local case. Once the period and the number of propagating modes are defined, numerical optimization is run to define the refractive index of each meta-atom to maximize the coupling in the desired direction. Fig.~\ref{fig:6_Non_local_responses} shows two configurations of strong non-local metasurfaces, an example of this design approach with very high energy efficiency. Fig.~\ref{fig:6_Non_local_responses}(a) represents the magnetic field of a metasurface designed with six elements per meta-cell that transmits the incoming plane waves at $45^{\circ}$ into the fourth diffraction mode in transmission, which makes an angle of $\theta_t=-45^{\circ}$ with $98.9\%$ energy efficiency. Alternatively, the same design approach can be used to control the reflection. As an example, Fig.~\ref{fig:6_Non_local_responses}(b) shows a configuration that produces retroreflection, $\theta_r=-45^{\circ}$, coupling the energy into the third Floquet mode with $91.2\%$ efficiency. 

The period of these two structures is not the same as the theoretical period dictated by a linear phase gradient. As it can be extracted from Eq.~(\ref{eq:general_diffraction}), when $k_{t}^{\rm(t,r)}=-k_{t}^{\rm(i)}$ with $k_{t}$ being the tangential wave number of the propagating waves, the total period defined by the linear phase gradient is $D=\lambda/(2\sin\theta_{\rm i})$~\cite{asadchy2017flat}. However, the cases refractive indexes for the PCMs layers in Fig.~\ref{fig:6_Non_local_responses} do not induce such phase profile. Moreover, each meta-atom by itself is designed to work as a Huygens source, Fig.~\ref{fig:3_Phase_and_transmittance}, therefore retroreflection, Fig.~\ref{fig:6_Non_local_responses}(b), is not even a logical EM response from a local response analysis.

\section{Electromagnetic coupling tunability}

In the previous section, we showed the control over the propagation properties of the proposed structure when exploiting electric and magnetic responses. Additional control of the optical response is possible making use of the bianisotropy effect. Bianisotropy refers to the induced electric/magnetic response produced by a magnetic/electric incident excitation. 
Depending on the specific kind of bianisotropy, the asymmetry may occur in reflection, transmission, or both~\cite{yazdi_bianisotropic_2015,alaee_all-dielectric_2015,radi_total_2013,radi_thin_2015,radi_one-way_2014} and can lead to reciprocal and non-reciprocal responses~\cite{asadchy2018bianisotropic,Mirmoosa_Polarizabilities_2014}.

\begin{figure}[t]
    \centering
\includegraphics[width=0.9\columnwidth]{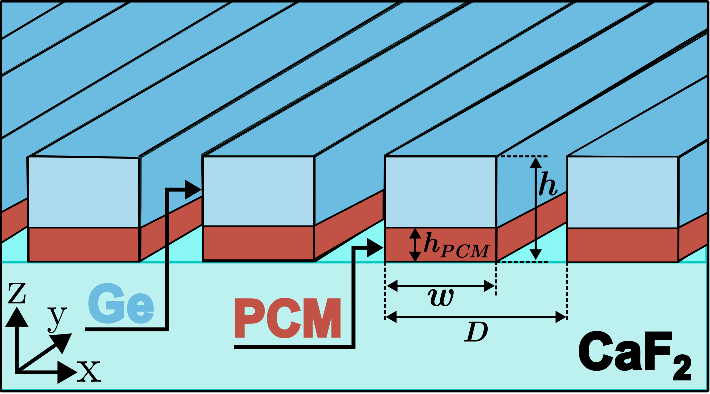}
    \caption{Metasurface representation and geometrical parameters. Infinite bars in the $y$-direction with rectangular cross-section defined by its height, $h$, and width $w$. The periodicity is defined as $D$. The PCM slab is introduced at the base of the bars and has height $h_{\rm PCM}$. A substrate of calcium difluoride is used to sustain the structure.}
    \label{fig:8_graphical_geometry_assymetric}
\end{figure}

Generally, bianisotropy inherently causes an asymmetry in the metasurface response if illuminated from different directions. In this section, we show that by integrating PCM in the meta-atoms, the EM coupling can be tuned, and the asymmetry of the response can be controlled. With the configuration presented in Fig.~\ref{fig:6_Non_local_responses}, mirror symmetry with respect to the z-axis is broken in the resonators. This last statement will be valid for our implementation, which will be discussed later, for every value of $n_{\rm PCM}$, as the top slab losses will differ from the bottom slab made out of PCM. The homogenization model used for analyzing the response of bianisotropic metasurfaces is ruled by the following constitutive relations~\cite{albooyeh_classification_2023-1}:
\begin{align}
\mathbf{P}=\frac{\overline{\overline{\alpha}}^{\mathrm{ee}}}{S} \cdot \mathbf{E}^{\mathrm{i}}+\frac{\overline{\overline{\alpha}}^{\mathrm{em}}}{S} \cdot \mathbf{H}^{\mathrm{i}},\\
\mathbf{M}=\frac{\overline{\overline{\alpha}}^{\mathrm{me}}}{S} \cdot \mathbf{E}^{\mathrm{i}}+\frac{\overline{\overline{\alpha}}^{\mathrm{mm}}}{S} \cdot \mathbf{H}^{\mathrm{i}}.
\end{align}
Where $\mathbf{E}^{\mathrm{i}}$ and $\mathbf{H}^{\mathrm{i}}$ are the incident electric and magnetic fields, respectively, $\mathbf{P}$ and $\mathbf{M}$ are the induced electric and magnetic surface polarization densities, respectively. The constitutive collective polarizabilities are the second-rank tensors  $\overline{\overline{\alpha}}^{\mathrm{ee}}$, for the electric coupling, $\overline{\overline{\alpha}}^{\mathrm{mm}}$, for the magnetic, $\overline{\overline{\alpha}}^{\mathrm{em}}$, for the magnetoelectric and $\overline{\overline{\alpha}}^{\mathrm{me}}$ for the EM. Here, $S$ stands for the unit cell surface. 

In the case of the structure shown in Fig.~\ref{fig:6_Non_local_responses}, the response can be modeled by using only a representation in cartesian components of the tangential part of polarizability tensors, as the metasurface is normally illuminated and the induced moments are going to be restricted to the $x$-$y$ plane. Furthermore, the geometry presented is non-chiral and is made out of an isotropic material, and the lattice only presents $\rm C_2$ symmetry, ensuring no cross-polarization conversion on a linear basis. In the following, the notation used is going to have the form $t_{ij}^{k}$ for transmission parameters or $r_{ij}^{k}$ for reflection parameters, where subscripts $i,j=\{+,-\}$ indicate the direction of propagation $+z$ or $-z$, respectively, and superscripts $k=\{x,y\}$ refer to the polarization of the incoming and outcoming field. Due to the discussed symmetry restrictions, the scattering matrix has some of its elements vanishing, $t_{xy}^{\pm}=t_{yx}^{\mp}=0$ and $r_{xy}^{\pm}=r_{yx}^{\pm}=0$. Moreover, due to reciprocity, $t_{xx}^+=t_{xx}^-$ and $t_{yy}^+=t_{yy}^-$. Therefore, only six parameters are independent in the scattering response problem: $t_{xx}$, $t_{yy}$, $r_{xx}^+$, $r_{xx}^-$, $r_{yy}^+$, and $r_{yy}^-$. Notice how, for the transmission coefficients, we have dropped the superscripts as reciprocity makes the direction of propagation irrelevant.

We aim to create a metasurface with a response that can be switched from transparent to one-way absorption (i.e., one side absorbs the incoming energy, and the other side acts as a reflector). Taking into account the restrictions in scattering parameter value and focusing on the $x$-polarized incident wave response, absorbance is defined for each side, or propagation direction, as  
\begin{equation}\label{eq:absorption}
    \mathrm{A}_{x}^{\pm}=1-|t_{xx}|^2-|r_{xx}^{\pm}|^2.
\end{equation}
The scattering parameter expressions in terms of the polarizabilities are~\cite{albooyeh_classification_2023-1}
\begin{equation}\label{eq:transmission}
t_{x x}^{ \pm}=1-\frac{j \omega}{2 S}\left(\eta \alpha_{x x}^{\mathrm{ee}}+\frac{\alpha_{y y}^{\mathrm{mm}}}{\eta}\right),
\end{equation}
\begin{equation}\label{eq:reflexion}
r_{x x}^{ \pm}=-\frac{j \omega}{2 S}\left(\eta \alpha_{x x}^{\mathrm{ee}}-\frac{\alpha_{y y}^{\mathrm{mm}}}{\eta} \mp 2 \alpha_{x y}^{\mathrm{em}}\right),
\end{equation}
where $\omega$ is the angular frequency, $\eta$ is the characteristic impedance of the host medium, and the polarizabilities subscripts indicate the matrix element of the representation of each polarizability tensor. Introducing Eq.~(\ref{eq:transmission}) and Eq.~(\ref{eq:reflexion}) into Eq.~(\ref{eq:absorption}) we obtain
\begin{equation}\label{eq:Absorption_total}
\begin{split}
   A_x^{\pm} = \frac{1}{\eta} ( \eta+\dfrac{\omega^2}{4S^2}(\alpha_{yy}^{\mathrm{mm}}-\eta( \alpha_{xx}^{\mathrm{ee}}\eta \mp 2\alpha_{xy}^{\mathrm{em}}))\\
  (\mp2{\alpha_{xy}^{\mathrm{em}}}^{\ast}+(-\alpha_{yy}^{\mathrm{mm}}/\eta+\alpha_{xx}^{\mathrm{ee}}\eta)^{\ast}-\\
 (i\eta+\dfrac{\omega}{2S}(\alpha_{yy}^{\mathrm{mm}}+\alpha_{xx}^{\mathrm{ee}}\eta^2))(-i+\dfrac{\omega}{2S}(\alpha_{yy}^{\mathrm{mm}}/\eta+\alpha_{xx}^{\mathrm{ee}})^{\ast})) ),
\end{split}  
\end{equation}
with $^\ast$ being the complex conjugation. Eq.~\ref{eq:Absorption_total} expression highlights that the difference in absorption from both sides is a direct cause of the EM coupling. However, it is not clear what is needed to obtain a high contrast, as the expression is too complex. To clarify the study, we define the contrast between absorptions for x-polarized light as 
\begin{equation}\label{eq:Delta_absorption_def}
    \Delta\mathrm{A}_x=\mathrm{A}_x^+-\mathrm{A}_x^-=|r_{xx}^{-}|^2-|r_{xx}^{+}|^2, 
\end{equation}
where the restrictions for the scattering parameters of the problem have been used to simplify the right side of the equation. Using this definition, Eq.~(\ref{eq:Delta_absorption_def}), and Eq.~(\ref{eq:reflexion}), one can show that
\begin{equation}\label{eq:Delta_absorption}
  \Delta \mathrm{A}_x = \dfrac{2\omega^2}{S^2}\Re\left[ {\alpha_{xy}^{\mathrm{em}}} \left({\eta\alpha_{xx}^{\mathrm{ee}}-\dfrac{1}{\eta}\alpha_{yy}^{\mathrm{mm}} }\right)^{\ast}\right],
\end{equation}

where $\Re$ stands for the real part. 

\begin{figure}[t]
    \centering
    \includegraphics[width=1\columnwidth]{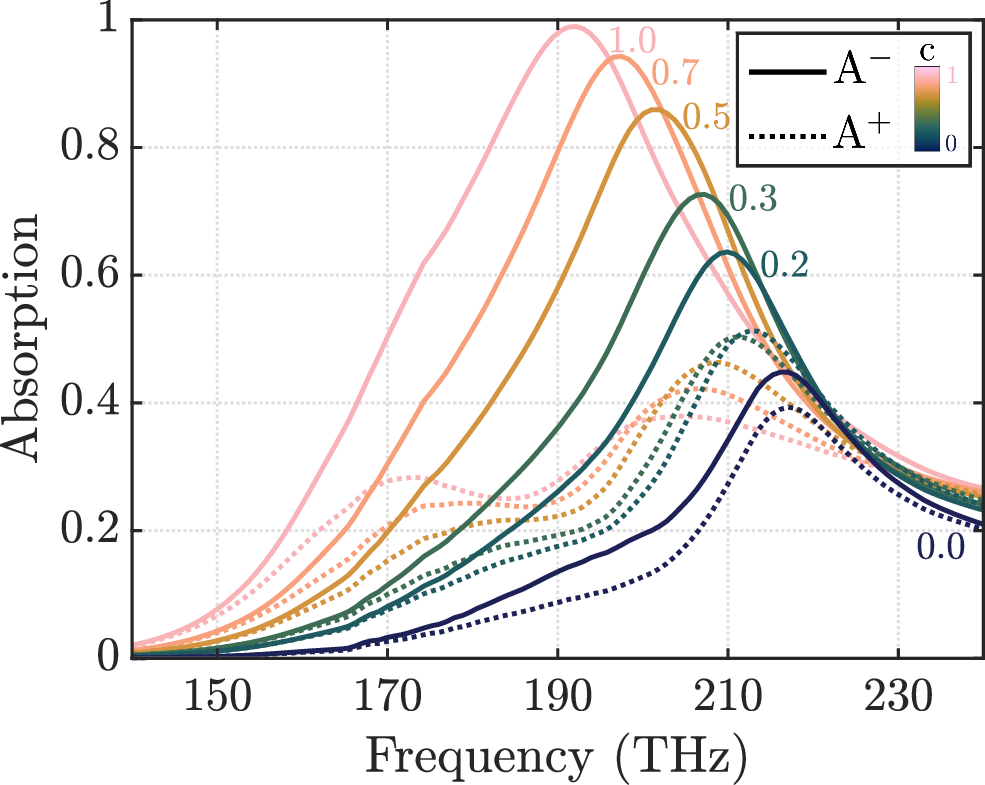}
    \caption{Absorption for several values of crystallization, $c$, obtained with normal incidence $x$-polarized plane wave illumination. Solid and dashed lines represent absorption for $-z$ and $+z$, respectively. The geometrical parameters used are $D=545.3$ nm, $\omega=210,2$ nm, $h=441.3$ nm, $h_{\rm PCM}=115.3$ nm.}
    \label{fig:9_Absorption_vs_crist}
\end{figure}

This result clearly shows that to achieve contrast, it is necessary, but not sufficient, to have EM coupling. In addition, Eq.~(\ref{eq:Delta_absorption}) proves that the bigger difference, for a given $\alpha_{xy}^{\mathrm{em}}$, between real and imaginary parts of $\alpha_{xx}^{\mathrm{ee}}$ and $\alpha_{yy}^{\mathrm{mm}}$ enlarges the absolute value of $\Delta\mathrm{A}_x$. It is known that by definition $\Delta\mathrm{A}_x$ is upper and lower bounded by 1 and -1, respectively. If Eq.~(\ref{eq:Delta_absorption}) is forced to be equal to 1, it imposes only one restriction to the six independent parameters in it, as the three polarizabilities are complex numbers. Nevertheless, Eq.~(\ref{eq:Delta_absorption}) tells us enough information about the path to follow. At the desired working frequency, a resonant state with EM coupling and a different response for magnetic and electric polarizabilities is necessary.

In our implementation, we choose $\rm{Ge_{2}Sb_{2}Te_{5}}$, a germanium antimony telluride (GST) composition, as our PCM. The characterization data for the material can be found in Ref.~\cite{frantz_optical_2023}. For the substrate, $\mathrm{CaF_2}$ is selected for its low refraction index at near-infrared frequencies considered $n_{\mathrm{CaF_2}}=1.42$. The refractive index of Germanium is modeled with the data shown in~\cite{amotchkina_characterization_2020}. Notice that in the near-infrared and visible regime, Germanium has non-negligible losses.

The results for the absorption of the metasurface, when illuminated by $x$-polarized plane waves at normal incidence, are shown in Fig.~\ref{fig:8_graphical_geometry_assymetric}. The crystallization state, $c$, is represented with different colors to show how, by tuning this magnitude, using heating devices or optical pumping, the absorption from one side of the metasurface can be changed from $9.7\%$ to $98.6\%$ at 193 THz while for illumination from the opposite side, it changes from $15.2\%$ to $31.0\%$. Therefore, at that frequency, the value of the contrast between each side's absorptions reaches $\Delta A_x=-0.676$ for $c=1$. The parameter $c$ ranges from 0 to 1 and accounts for the fraction of material that has been crystallized.

\section{Conclusions}

This work presents a simple design for a reconfigurable all-dielectric metasurface formed by nanobars that combine high-dielectric materials with PCMs. The fundamental feature of the proposed platform is its capability to fully control the EM response: electric, magnetic, and EM coupling by changing the refractive index of the PCMs inside the resonators. Empowered by this broad control, two applications are shown: beam steering and switchable asymmetric absorption with out-of-band transparency.

By harnessing the principles of a Huygens source, this configuration offers practical applications in designing metasurfaces that can alter the direction of transmitted waves. This is achieved through a local approach, implementing a linear phase gradient along the metasurface. The ability to control the number of resonators used to create the gradient allows for beam steering. Moreover, the metasurface can be programmed to utilize auxiliary evanescent fields, creating strong non-local effects that can tailor the energy to different diffraction modes, thereby surpassing the limitations of local designs.

Finally, we show that one can control EM coupling in the structure by breaking the symmetry of the structure in the $z$-direction by placing the layer of PCM asymmetrically. With this asymmetric configuration, we can obtain different light absorption when the metasurface is illuminated from opposite sides. As an example, we show how to design the metasurface to obtain absorption asymmetry. Furthermore, controlling the state of the PCM in the structure enables the possibility of switching from transparent to one-way absorption at optical frequencies.

\begin{acknowledgments}
L.M.M-E. acknowledges Universitat Polit\`ecnica de Val\`encia (PAID-01-23). A.D.-R. acknowledges Beatriz Galindo excellence grant (grant No.BG-00024) and the Spanish National Research Council (grant No.PID2021-128442NA-I00). Color palettes and gradients used in the paper were extracted from Ref.~\cite{crameri_misuse_2020}.
\end{acknowledgments}

\bibliography{main.bib}

\end{document}